%% file: main.tex
\documentclass[sigconf]{acmart}

\usepackage[utf8]{inputenc}
\usepackage{xspace}
\usepackage{url}
\usepackage{booktabs}
\usepackage{graphicx}
\usepackage{multirow}
\usepackage{color, soul}
\newcommand{\ie}{\textit{i.e., \xspace}}
\newcommand{\eg}{\textit{e.g., \xspace}}
\newcommand{\etal}{\textit{et al. \xspace}}
\newcommand{\toolname}{\textsc{AntiCopyPaster}\xspace}

\makeatletter
\newcommand\footnoteref[1]{\protected@xdef\@thefnmark{\ref{#1}}\@footnotemark}
\makeatother

\ccsdesc[500]{Software and its engineering~Software evolution}
\ccsdesc[500]{Software and its engineering~Maintaining software}

\keywords{refactoring, software quality, machine learning}

\makeatletter
\newcommand{\linebreakand}{%
  \end{@IEEEauthorhalign}
  \hfill\mbox{}\par
  \mbox{}\hfill\begin{@IEEEauthorhalign}
}
\makeatother

\title[\toolname: Extracting Code Duplicates As Soon As They Are Introduced in the IDE]{\toolname: Extracting Code Duplicates \\ As Soon As They Are Introduced in the IDE}

\author{Eman Abdullah AlOmar$^{*}$}
\affiliation{
    \institution{Stevens Institute of Technology}
    \city{Hoboken}
    \country{United States}
}
\email{ealomar@stevens.edu}

\author{Anton Ivanov$^{*}$}
\affiliation{
    \institution{HSE University}
    \city{Moscow}
    \country{Russia}
}
\email{apivanov_1@edu.hse.ru}

\author{Zarina Kurbatova**}
\affiliation{
    \institution{Huawei}
    \city{Saint Petersburg}
    \country{Russia}
}
\email{zarina.kurbatova@huawei.com}

\author{Yaroslav Golubev}
\affiliation{
    \institution{JetBrains Research}
    \city{Belgrade}
    \country{Serbia}
}
\email{yaroslav.golubev@jetbrains.com}

\author{Mohamed Wiem Mkaouer}
\affiliation{
    \institution{Rochester Institute of Technology}
    \city{Rochester}
    \country{United States}
}
\email{mwmvse@rit.edu}

\author{Ali Ouni}
\affiliation{
    \institution{ETS Montreal, University of Quebec}
    \city{Montreal, Quebec}
    \country{Canada}
}
\email{ali.ouni@etsmtl.ca}

\author{Timofey Bryksin}
\affiliation{
    \institution{JetBrains Research}
    \city{Limassol}
    \country{Cyprus}
}
\email{timofey.bryksin@jetbrains.com}

\author{Le Nguyen}
\affiliation{
    \institution{Rochester Institute of Technology}
    \city{Rochester}
    \country{United States}
}
\email{ln8378@rit.edu}

\author{Amit Kini}
\affiliation{
    \institution{Rochester Institute of Technology}
    \city{Rochester}
    \country{United States}
}
\email{ak3328@rit.edu}

\author{Aditya Thakur}
\affiliation{
    \institution{Rochester Institute of Technology}
    \city{Rochester}
    \country{United States}
}
\email{at4415@rit.edu}
\thanks{* These authors contributed equally.\\ ** The work was carried out when the author worked at JetBrains Research}

\begin{document}

\begin{abstract}

We developed a plugin for IntelliJ IDEA called \toolname, which tracks the pasting of code fragments inside the IDE and suggests the appropriate \textit{Extract Method} refactoring to combat the propagation of duplicates. Unlike the existing approaches, our tool is integrated with the developer's workflow, and pro-actively recommends refactorings. 
Since not all code fragments need to be extracted, we develop a classification model to make this decision. When a developer copies and pastes a code fragment, the plugin searches for duplicates in the currently opened file, waits for a short period of time to allow the developer to edit the code, and finally inferences the refactoring decision based on a number of features.

Our experimental study on a large dataset of 18,942 code fragments mined from 13 Apache projects shows that \toolname correctly recommends \textit{Extract Method} refactorings with an F-score of 0.82. Furthermore, our survey of 59 developers reflects their satisfaction with the developed plugin's operation.
The plugin and its source code are publicly available on GitHub at 
\url{https://github.com/JetBrains-Research/anti-copy-paster}. The demonstration video can be found on YouTube: \url{https://youtu.be/_wwHg-qFjJY}.

\end{abstract}
\maketitle
\input{sections/01-introduction}
\input{sections/02-approach}
\input{sections/03-implementation}
\input{sections/04-evaluation}
\input{sections/05-conclusion}

\bibliographystyle{ACM-Reference-Format}
\bibliography{cites}

\end{document}

%% file: sections/01-introduction.tex
\section{Introduction}

\begin{figure*}
  \centering
  \includegraphics[width=.8\textwidth]{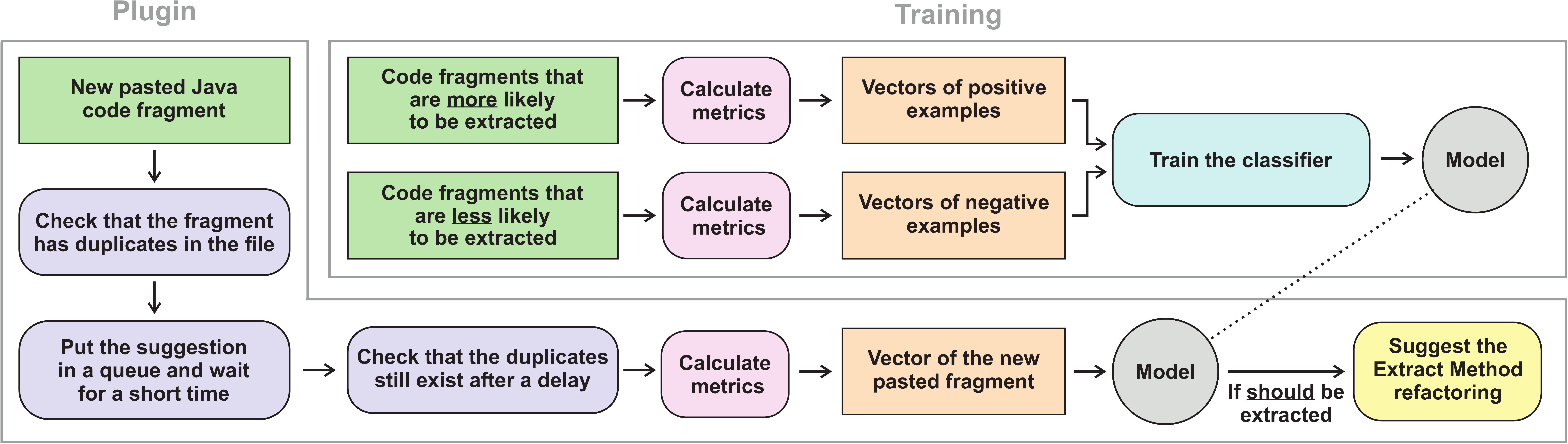}
  \vspace{-0.3cm}
  \caption{The pipeline of \toolname. \textit{Top}: training the model, \textit{bottom}: using the plugin.}
  \label{fig:pipeline}
  \vspace{-0.3cm}
 \end{figure*}
 
Copying and pasting code constitute an intuitive practice when writing source code. The resulting \textit{duplicate code} helps developers implement similar functionalities and optimize the working process. Prior research showed that a significant percentage of computer code consists of duplicate code~\cite{lopes2017dejavu,golubev2020study}, \textit{i.e.}, fragments of code that are similar or exactly the same.
 
Recent studies have shown that duplicate code brings its own challenges, including bug propagation in duplicated fragments~\cite{thongtanunam2019will}. Hence, removing duplicate code via refactoring has become a common emerging solution~\cite{fanta1999removing}. Refactoring duplicate code consists in taking a code fragment and moving it to create a new method, while replacing all instances of that fragment with a call to this newly created method. This refactoring is known as \textit{Extract Method} \cite{fowler2018refactoring}. 

Despite the existence of several studies recommending the refactoring of code duplication~\cite{yoshida2019proactive,alcocer2020improving,hotta2010duplicate}, their adoption is challenged by the need to exhaustively search the entire code base to recommend proper \textit{Extract Method} refactorings. That is, the whole source code is used as input to list various \textit{Extract Method} refactorings for developers to apply. Such a solution makes a strong assumption that developers have the expertise of the entire system and the separate time to consider their options. 

To cope with these challenges, this paper aims to support programmers with \textit{just-in-time} refactoring of duplicate code by designing \toolname, an IntelliJ IDEA\footnote{IntelliJ IDEA: \url{https://www.jetbrains.com/idea/}} plugin that refactors duplicate code as soon as it is introduced. \toolname is automatically launched when a duplicate piece of code is pasted and is not edited for some time, by displaying a background pop-up notification, alerting the developer of a potential \textit{Extract Method} refactoring. The developer can choose whether to click on the notification or ignore it, until it disappears after a few seconds. If the notification is clicked, the \textit{Extract Method} refactoring feature is called with the duplicate code as input, and a refactoring preview window is opened. The developer can then apply the refactoring and suggest a name to the newly created method.

The main advantage of our tool, in contrast with previous works, is the recommendation of refactoring a code fragment that is within the current context of development. However, not all duplicate code fragments need to be refactored, and the main challenge is to efficiently recommend refactoring only when the refactoring is \textit{worth it}, to avoid annoying developers with a recommendation whenever a source code is copied and pasted. The decision of whether the given duplicate code fragment should be extracted is defined as a binary classification problem. The duplicate fragment is parsed using the IDE's Program Structure Interface (PSI)\footnote{PSI: \url{https://plugins.jetbrains.com/docs/intellij/psi-files.html}} to generate its corresponding syntactic and semantic model. This model is used to calculate a set of 78 comprehensive structural and semantic metrics, previously used in various studies recommending the \textit{Extract Method} refactoring ~\cite{shahidi2022automated,tiwari2022identifying,silva2015jextract,haas2016deriving,aniche2020effectiveness,van2021data,ouni2016multi}. The metrics' values are given as input to a binary classifier, which was trained on a large dataset of 18,942 previously performed \textit{Extract Method} refactorings. 

We evaluate the \textit{correctness} of \toolname on the same dataset of 18,942 code fragments mined from 13 mature Apache projects, which shows that the used CNN model achieves an F-measure of 0.82. Also, we evaluate the \textit{usefulness} of \toolname through a survey with 59 participants. The survey responses are promising, as the majority of participants are satisfied with the recommendations of \toolname.

%% file: sections/02-approach.tex
\section{Approach}
\label{sec:approach}

In a nutshell, the goal of \toolname is to automatically provide \textit{just-in-time} recommendations of \textit{Extract Method} refactoring opportunities as soon as duplicate code is introduced in the opened file in the IDE. Our tool takes various semantic and syntactic code metrics as input and makes a binary decision on whether the code fragment has to be extracted.  The overall framework of our approach is depicted in Figure~\ref{fig:pipeline}. 
The tool and the dataset can be found on the project's GitHub page. 

\textbf{Data Collection}. Our first step consists of selecting 13 mature projects from the Apache Software Foundation,\footnote{Apache projects on GitHub: \url{https://github.com/apache}} which are popular open-source Java projects hosted on GitHub \cite{bavota2015apache}. These curated projects were selected with respect to both project size and activity.

\textbf{Refactoring Detection}. To extract the entire refactoring history of each project, we used RefactoringMiner v2.0,\footnote{RefactoringMiner: \url{https://github.com/tsantalis/RefactoringMiner}} a widely-used refactoring detection tool introduced by Tsantalis~\etal~\cite{tsantalis2020refactoringminer}. We identify methods that underwent an \textit{Extract Method} refactoring (\ie positive examples) using RefactoringMiner. In total, the tool mined 9,471 cases of \textit{Extract Method} refactorings. Specifically, we discovered \textit{Extract Method} refactorings, then traversed the history to the previous commit and took the code fragment that had been extracted. This allowed us to detect fragments that are \textit{worth} to be extracted, since they were extracted in mature projects. These refactorings are not necessarily only applied in the context of duplicate code, and thus our model learns from various contexts (\eg splitting long methods). 
To collect the negative samples, we start with selecting all sequences of statements that are eligible to be extracted. Then, they are ranked according to a special scoring formula inspired by the work of Haas and Hummel~\cite{haas2016deriving}. While their approach is aimed to find fragments that \textit{should} be extracted, we use the bottom 95\% of the ranked list to find code fragments that are \textit{less likely} to be extracted. Then, to create a balanced dataset, we sampled 9,471 fragments.

\textbf{Code Metrics Selection}. After collecting positive and negative examples, we characterize them through various metrics. The goal of selecting metrics is to identify patterns in their values to allow distinguishing between the two classes of fragments. To do so, we gathered all the metrics that have been extensively used in previous studies~\cite{aniche2020effectiveness,haas2016deriving} and then removed all the redundant metrics to avoid generating features with similar values. In total, we selected 78 metrics that can be related to the code fragments, enclosing methods, and coupling.

\textbf{Model Training}. We define the detection of an \textit{Extract Method} opportunity as a binary classification problem. Our intended model takes a set of metrics as input, and uses them as features to learn patterns in their values that distinguish between duplicate code fragments that are more likely and less likely to be extracted. Since the input corresponds to 78 metrics, we chose to rely on Convolutional Neural Networks (CNNs) 
for building our model. 

%% file: sections/03-implementation.tex
\section{Tool implementation}
In this section, we describe the specific implementation of our plugin for IntelliJ IDEA. The plugin consists of four main components. 

\textbf{Duplicate Detector}. To detect duplicates, we use bag-of-words token-based clone detection~\cite{sajnani2016sourcerercc}. This code similarity-based approach takes a given code fragment as input, then parses all methods inside the same file, so that each method is represented as tokens. The next step is to compute the similarity between the code fragment and methods via their abstracted token representation. This approach can detect an exact match, \ie when the code fragment is a substring of the method body. The bag-of-tokens similarity also takes into account minor changes in the pasted fragment, such as reordering the sequence of code, or renaming an identifier.

Since it is possible that a code fragment will be significantly edited soon after it is pasted, in order to avoid the immediate flagging of the pasted code as duplicate, and potentially interfering with the developer's flow, we implement a \textit{delay} and place the pasted code fragment in a queue. Then, two sanity checks are executed: we check whether the pasted fragment is Java code and whether it constitutes a correct syntactic statement. To do that, the plugin tries to build a PSI tree of the fragment. A PSI (Program Structure Interface) tree is a concrete syntax tree that is used in the IntelliJ Platform to represent the structure of code. If a PSI tree can be built and represents a valid statement, and if the duplicates still remain after the delay, the code fragment is passed to the \textit{Code Analyzer}.

\textbf{Code Analyzer.} This component takes the duplicate fragment as input and uses its PSI representation to calculate the 78 metrics discussed above. The code fragment, with its corresponding vector of metrics, constitute the input to the \textit{Method Extractor}.

\textbf{Method Extractor.} This component takes as input the vector of metrics, and feeds it to the pre-trained model in order to make the binary decision of whether this code fragment is similar to the ones that have been previously refactored in real projects. If the classifier confirms the refactoring, then \textit{Refactoring Launcher} is called.

\textbf{Refactoring Launcher.} This component starts with checking if the pasted code fragment could be extracted into a separate method without any compilation errors. If all checks pass, a notification is then enabled to appear in the bottom right corner of the editor, informing the developer that an \textit{Extract Method} refactoring is recommended. If the user responds to the tip, \textit{Refactoring Launcher} passes the duplicate fragment as an input to the IDE's built-in Extract Method API, and initiates the preview window. The user has the choice to either confirm the refactoring, while renaming the newly extracted method, or cancel the entire process.

\begin{figure}
  \centering
  \includegraphics[width=.9\columnwidth]{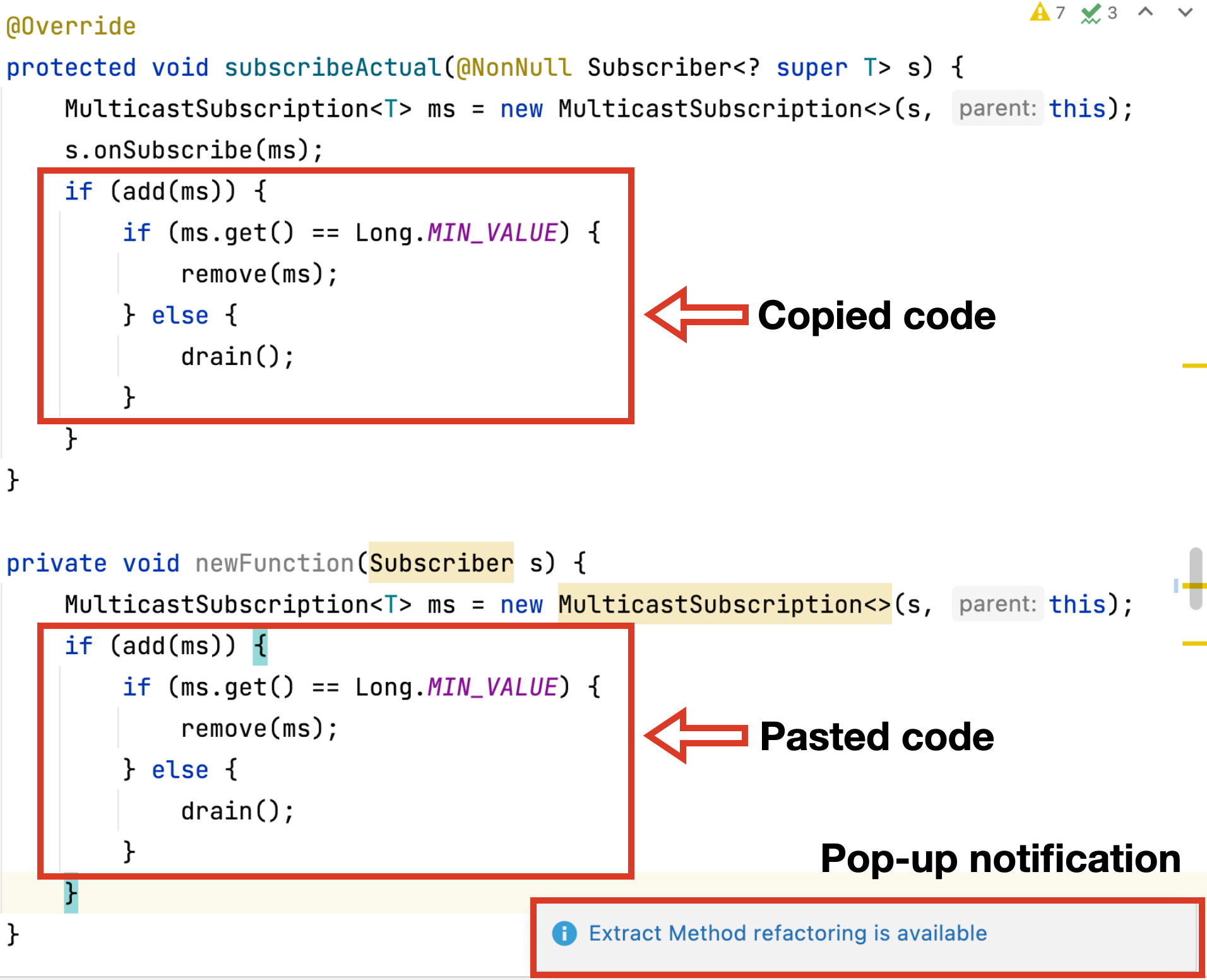}
  \caption{\textit{Extract Method} refactoring opportunity. }
  \label{fig:example}
  \vspace{-0.3cm}
 \end{figure}

We further illustrate \toolname in Figure \ref{fig:example}, showing an example of a duplicate piece of code  pasted and not edited for some time, and a pop-up notification appearing at the bottom of the IDE, alerting the developer of a potential \textit{Extract Method}.

%% file: sections/04-evaluation.tex
\section{Evaluation}

\subsection{Correctness}

We test the ability of our Convolutional Neural Network (CNN) to accurately recommend \textit{Extract Method} refactoring opportunities. Further, we compare the performance of our CNN model with four machine learning classifiers: Random Forest (RF), Support Vector Machine (SVM), Naive Bayes (NB), and Logistic Regression (LR).   
The selection of these ML classifiers was due to the fact that their performance was competitive in similar binary classification problems~\cite{aniche2020effectiveness,alomar2021we,alomar2021toward,Levin:2017:BAC:3127005.3127016}.
In order to evaluate the performance of the algorithms, we use out-of-sample bootstrap validation since this validation technique yields the best balance between the bias and variance in comparison to single-repetition holdout validation~\cite{tantithamthavorn2016empirical}.

The comparison between the classification algorithms is reported in Table \ref{Table:binary_classifiers}. 
Based on our findings, the F-measure of CNN is 82\%, higher than its competitors RF, SVM, NB, and LR, achieving 81\%, 76\%, 56\%, and 71\%, respectively. We conjecture that a proper conveyance of the semantics behind the source code would have required complex feature engineering using neural network classification strategy rather than traditional machine learning algorithms. This observation has been also supported by previous studies that utilized deep learning to source code analysis~\cite{zampetti2020automatically,tufano2019empirical}.

Despite the fact that there is no model that outperforms all the others in both precision and recall, the choice of the model can become the decision of the practitioner who is adopting the tool. 
Additionally, it is important to consider the practicality of using different models. From this standpoint, the trained CNN is smaller than a Random Forest and loads faster into the memory. At the same time, our particular implementation of CNN required the use of the TensorFlow framework, which added a lot of overhead to the plugin. In future work, we plan to consider other potential libraries and frameworks for inferencing ML models.

\input{tables/results}

\vspace{-0.2cm}
\subsection{Usefulness}

To evaluate the usefulness of \toolname, we performed an external validation by involving 96 participants from the Rochester Institute of Technology, Stevens Institute of Technology, and ETS Montreal. 
In total, 59 developers participated in the survey (yielding a response rate of 61.4\%, which is considered high for software engineering research \cite{smith2013improving}), and 39 of them executed the plugin and tested it thoroughly.

\begin{figure}[t]
\centering
\includegraphics[width=.75\columnwidth]{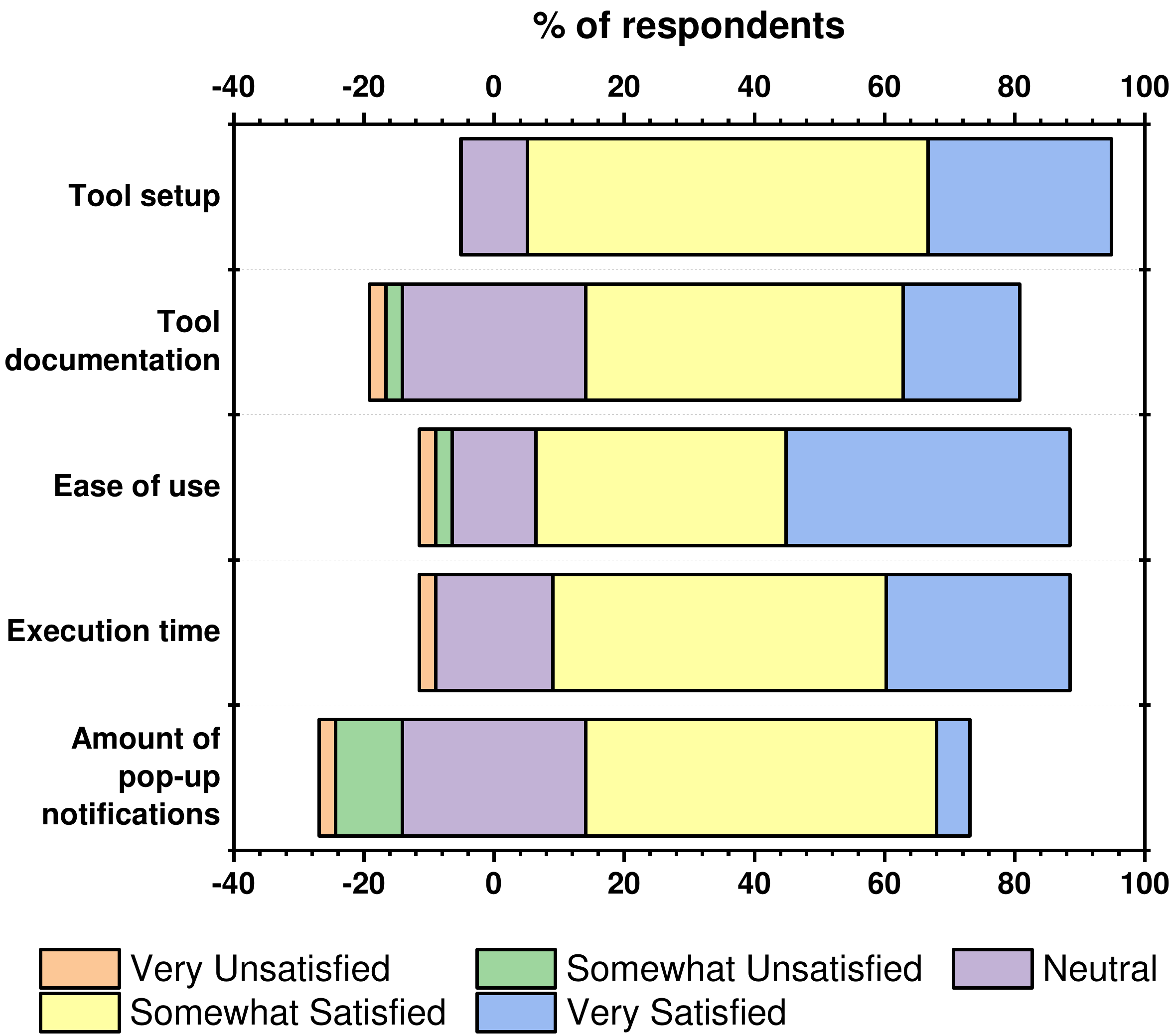}
\caption{Participants' satisfaction with various  aspects of the \toolname tool.}
\label{fig:likertscale}
\vspace{-0.3cm}
\end{figure}

Figure~\ref{fig:likertscale} depicts an overview of their answers. With respect to the tool setup, most of the respondents reported that they are satisfied with the tool. Regarding the tool documentation, the majority of the respondents agreed that the documentation is useful; only a couple of participants were unsatisfied. For the ease of use aspect, a larger group was satisfied. Several participants found that the tool is not easy to use, so we will work on improving the usability of the tool.  Concerning the execution time, most of the participants were happy with it. For the amount of pop-up notification, the majority of respondents agreed that the amount of pop-up notifications is acceptable.
There were also a few participants who were not happy with the amount of pop-up notification, and we are planning on improving this aspect of the tool in the future.

%% file: tables/results.tex
\begin{table}[]
\begin{center}
\caption{The performance of different classifiers.}
\label{Table:binary_classifiers}
\vspace{-0.3cm}
\scriptsize
\begin{tabular}{lcccc}\hline
\toprule
\bfseries Classifier & \bfseries Precision & \bfseries Recall & \bfseries F-measure & \bfseries PR-AUC  \\
\midrule
Random Forest & \textbf{0.83} & 0.78 & 0.81 & \textbf{0.86}\\
Support Vector Machine & 0.78 & 0.74 & 0.76 & \textbf{0.86}\\
Naive Bayes & 0.72 & 0.46 & 0.56 & 0.72\\
Logistic Regression & 0.73 & 0.70 & 0.71& 0.79\\
\textbf{Convolutional Neural Network} & 0.82 & \textbf{0.82} & \textbf{0.82} & \textbf{0.86}
 \\
\bottomrule
\end{tabular}
\end{center}
\vspace{-0.4cm}
\end{table}

%% file: sections/05-conclusion.tex
\section{Conclusion}

Recommending \textit{Extract Method} refactoring opportunities is critical to both the research community and industry. Despite the fact that numerous research works have used a number of ways to discover \textit{Extract Method} refactoring, advocating this refactoring type without interfering with developers' workflow has largely remained unexplored. In this study, we proposed \toolname as an IntelliJ IDEA plugin, and experimented with machine learning models in order to increase the adoption and usage of the \textit{Extract Method} refactoring while maintaining the workflow of a developer. Our findings show that machine learning models are efficient in identifying \textit{Extract Method} refactoring opportunities as soon as code duplicates are presented in the IDE, and that the \toolname tool was well received by developers.